# RELATIONSHIP BETWEEN EARTH-DIRECTED SOLAR ERUPTIONS AND MAGNETIC CLOUDS AT 1AU: A BRIEF REVIEW

VASYL YURCHYSHYN

*Big Bear Solar Observatory 40386 North Shore Lane, Big Bear City, CA 92314*

DURGESH TRIPATHI

*Department of applied Mathematics and Theoretical Physics, University of Cambridge, Wilberforce Road, Cambridge CB30WA, UK*

We review relationships between coronal mass ejections (CMEs), EIT post eruption arcades, and the coronal neutral line associated with global magnetic field and magnetic clouds near the Earth. Our previous findings indicate that the orientation of a halo CME elongation may correspond to the orientation of the underlying flux rope. Here we revisit these preliminary reports by comparing orientation angles of elongated LASCO CMEs, both full and partial halos, to the post eruption arcades. Based on 100 analysed events, it was found that the overwhelming majority of halo CMEs are elongated in the direction of the axial field of the post eruption arcades. Moreover, this conclusion also holds for partial halo CMEs as well as for events that originate further from the disk center. This suggests that the projection effect does not drastically change the appearance of full and partial halos and their images still bear reliable information about the underlying magnetic fields. We also compared orientations of the erupted fields near the Sun and in the interplanetary space and found that the local tilt of the coronal neutral line at 2.5 solar radii is well correlated with the magnetic cloud axis measured near the Earth. We suggest that the heliospheric magnetic fields significantly affect the propagating ejecta. Sometimes, the ejecta may even rotate so that its axis locally aligns itself with the heliospheric current sheet.

## 1. Introduction

Geomagnetic storms are major disturbances in the Earth's magnetosphere that occur when the interplanetary magnetic field turns southward and remains so for a prolonged period of time [1, 2, 3]. Reconnection between the southward-directed (relative to the ecliptic plane) component of the solar wind-carried magnetic field, $B_z$, and the northward-directed geomagnetic field can occur at the dayside magnetopause, resulting in the transfer of significant amounts of energy from the solar wind into the Earth's magnetosphere. During a typical storm, an enhanced ring current is created and the auroral electrojets in both hemispheres move (often dramatically) equatorward. This is evidenced by a decrease in the geomagnetic field at near-equatorial magnetic stations, and a large decrease in the *Dst* index that is currently compiled from the records from four selected stations [4].

All regions of geospace are affected by geomagnetic disturbances, and the operations of various technological systems can be impaired or even totally disrupted. For example, during large geomagnetic disturbances the rapidly changing electric currents in the ionosphere (regions within the moving electrojets as well as at distances from them), largely driven by the precipitation of charged particles from the magnetosphere, cause rapid changes in the magnetic field at Earth's surface. These changing fields in turn can induce enhanced electric currents to flow in the Earth's crust and mantle ("telluric currents"). These currents seek the highest conductivity paths, which are often long, e.g., grounded electric grid systems and communications cables. As another example, the enhanced disturbances in the ionosphere can produce serious scintillations in trans-ionospheric radio-wave propagation, even at GHz frequencies. Scintillation effects can be especially large in the equatorial regions at times (producing convective ionospheric storms, often referred to as sporadic E-events), but can be significant across all latitudes.

The occurrence of geomagnetic storms is well associated with Earth-directed coronal mass ejections (CMEs), which appear in coronagraph images as bright halos around the occulting disk [halo CME, 5]. CMEs are eruptions of the solar magnetic field and plasma into interplanetary space, which occur following a large-scale magnetic rearrangement in the solar atmosphere [6, 7, 8, 9, also see several excellent reviews in Space Science Rev., 2006, DOI: 10.1007/s11214-006-9027-8]. Studies of soft X-ray images of the solar corona indicate that eruptions are more likely to occur in large magnetic configurations that contain a bright sigmoidal structure [10], which is thought to be related to magnetic flux ropes [11, 12]. Potential magnetic field modelling [13, 14] suggests that the origin of CMEs is determined by the global structure of the solar magnetic field and is related to the helmet streamers that are sensitive to the evolution of active regions and/or filaments.

Depending upon the orientation of the magnetic field in the CME-originating region and the sign of magnetic helicity (twist), the Earth-directed CME may or may not have an intense negative $B_z$ field [15]. Hence, the origin of CMEs, the structure of their source regions and their signatures in the solar wind near the Earth are of fundamental interest in the physics of the Sun, space plasmas, and space-weather research.

This paper is organised as follows. In Section 2 we describe the relationship between CME source regions and interplanetary ejecta. Particular attention is paid to the magnetic structure of a CME, and how it corresponds to that of the associated post eruption arcades (PEAs). Section 3 compares magnetic orientations of halo CMEs, tilt of the main coronal neutral line near the eruption site and the directional parameters of the corresponding magnetic clouds. In Section 4 we briefly summarise the results and discuss their implications.

## 2. Relationship between magnetic fields of CME source regions and the corresponding interplanetary ejecta

Interplanetary counterparts of CMEs are called interplanetary CMEs (ICMEs) or interplanetary ejecta. The internal structure of an ICME is not trivial and may include an ICME body that is often preceded by a shock. The ICME body, in turn, may be observed as a complex ejection [16] or a magnetic cloud [MC, 17].

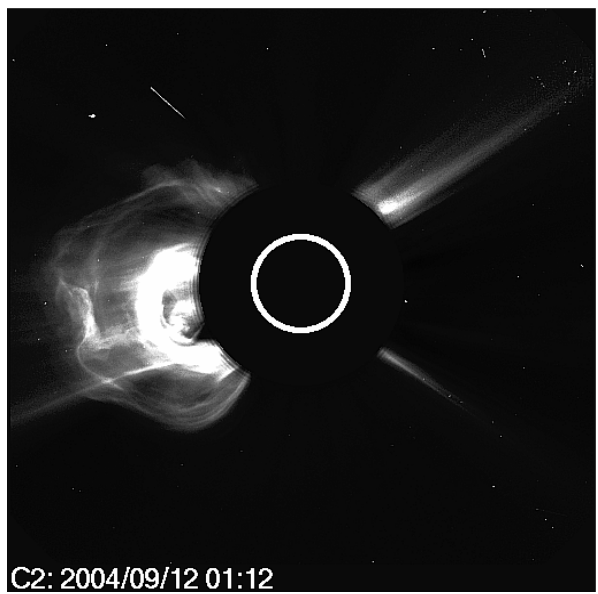

*Fig. 1. LASCO C2 image of a limb CME that erupted on 12 September 2004. This image represents an event with a distinct three part structure and suggests that, at least some CMEs can be described as a flux rope structures.*

While MCs are known to be the cause of the most severe geomagnetic storms [18], interplanetary shocks [19] and co-rotating interaction regions [9] are also quite geoeffective, although the storm associated with them are, on average, less intense than those associated with the MCs.

For a number of years researchers have been trying to establish a relationship between CMEs, their complex progenitors, and interplanetary ejecta [20]. Since complex ejecta represent unorganized and compound magnetic structures they were not subject of these studies. Instead, effort was focused on MCs that generally exhibit a magnetically organized geometry, which is thought to correspond globally to a curved flux rope as shown in Figure 1 [also see Figures 1 and 2 in 21, 22, 23] and can be related to various solar parameters such as magnetic field helicity, strength and direction [8]. In particular, significant research efforts have been devoted to finding whether the direction of the magnetic field and the sense of twist found in MCs correspond to those associated with CME source regions (which include both active regions and filaments).

### 2.1 Association between Ejecta and Post Eruption Arcades

According to various case and statistical studies many interplanetary ejecta maintain nearly the same orientation and twist as the source regions they are associated with [e.g., 12, 15, 22, 24, 25, 26]. This is in agreement with helicity conservation law established for laboratory plasmas. Several statistical studies, however, produced inconclusive results [27, 28]. Harra *et al.* [29] conducted a thorough case study of flaring activity and concluded that the same active region may produce interplanetary ejecta with different magnetic orientations. Therefore, the question about the magnetic connection between interplanetary ejecta and solar surface phenomena is still open. There are no well defined schemes to predict the magnetic field structure at 1AU based on solar surface measurements present at this moment.

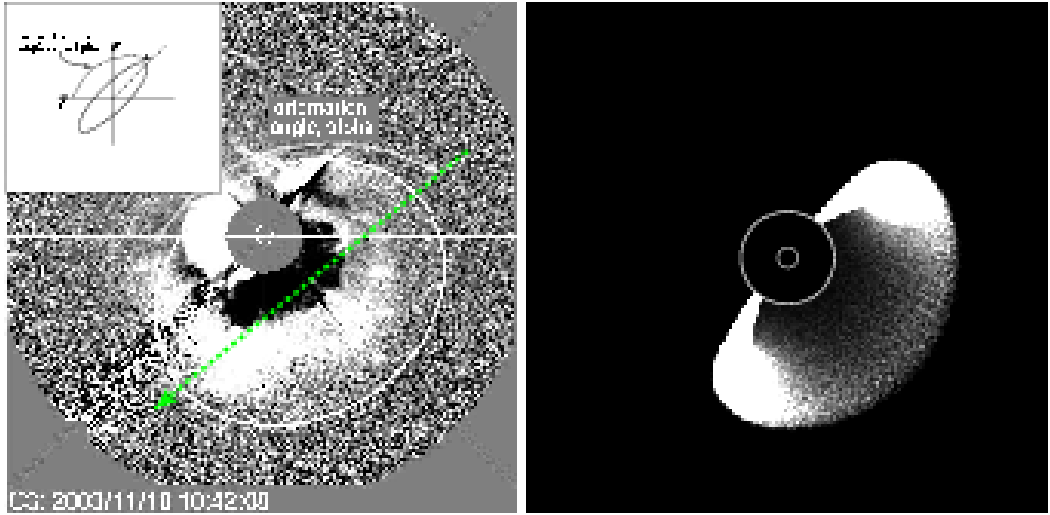

*Fig. 2. Left: LASCO C3 difference image for the 2003 November 18 eruption used to measure the CME elongation. The halo CME was measured at its outer edge at intervals of 45 degrees (lines). An ellipse was then fitted to the eight halo points and its tilt angle was determined in the clockwise direction from the y-axis that points eastward to the ellipse semi-major axis (see inset). Right: A synthetic "halo" CME image produced for the same event by the erupting flux rope model. The tilt of the synthetic elongated halo represents the orientation of the simulated flux rope. Comparison with the left panel shows that the tilts of those two halos are similar thus indicating the correspondence between the observed halo elongation and the magnetic axis of a flux rope.*

Here, we further explore this problem by utilising magnetic field information measured at various places between the solar surface and the Earth. In particular, we compare directional angles of halo CMEs, heliospheric current sheet (HCS), MCs, and EUV PEA, observed with Extreme Ultraviolet Imaging Telescope [EIT, 30] on board Solar and Heliospheric Observatories [31].

2.1.1. *Measurements of CME orientations from white light coronagraph images*

White light coronagraph data, such as provided by the Large Angle Spectroscopic COronagraph LASCO [32] instrument also on board the SOHO spacecraft, show that halo CMEs have various sizes and shapes. Many of them can be enveloped by an ellipse and can be fitted with a cone model [33, 34, 35, 36]. CMEs are also found to be highly structured three-dimensional features [37, 38] and they are thought to represent a flux rope [17, 38, 39, 40, 41, 42]. In particular, Krall & St. Cyr [38] showed that statistical parameters measured for observed CMEs, such as eccentricity and the axial aspect ratio, are in agreement with those obtained for a parameterized model flux rope.

The structured CMEs appear to be magnetically organised in the axial direction, which corresponds to the axis of a large-scale twisted flux rope [37]. At the same time, modelling of erupting flux ropes [43, 44, 45] shows that modelled halo CMEs appear to be elongated in the direction of the axial field of the underlying flux rope. Therefore, it is possible the ellipse-shaped appearance of halo CMEs may be related to their originating magnetic structure.

Yurchyshyn *et al.* [45, hereafter Paper I] studied elongation parameters of 25 halo CMEs and compared them to the corresponding parameters determined for MCs. The CME-MC events were selected from the Master Data Table compiled during a *Living With a Star* Coordinated Data Analysis Workshops [46, 47] and the list published in [48]. All orientation angles were determined in the geo-centric solar ecliptic coordinate system (GSE), where *y*-axis is in the ecliptic plane pointing towards dusk, *x*-axis is directed from the Earth towards the Sun and *z*-axis is pointed northward. The CME orientation angle (or tilt) was determined in Paper I by fitting an ellipse to an irregularly shaped "halo" around the C3 occulting disk (Figure 2) and measuring its tilt in the clockwise direction from the positive GSE *y*-axis to the ellipse semi-major axis. For each event, we measured the angles in 3-5 images depending upon their availability.

The final tilt angle listed in Table 1 of Paper I was calculated as the mean of angles determined from individual frames.

*2.1.2. Orientation of CME elongations versus direction of the axial field in magnetic clouds*

The main conclusion drawn from the comparison of CME elongations and MC orientations, discussed above, is that for about 64% of CME-MC events there is a good correspondence between their orientation angles. This supports the idea that the elongation of halo CMEs may indeed reflect the underlying structures of the erupting magnetic fields. If that is the case, these results also imply that the majority of interplanetary ejecta do not significantly change their orientations (less than 45 deg rotation) while traveling from the Sun to the near Earth environment. This was later confirmed by Zhao (2007, private communication). On the other hand, not-so-close-correspondence between CME and MC orientations seems to reflect the existing general confusion about the relationship between solar and MC fields: while for the majority of events magnetic fields at the Sun and 1AU seem to be in agreement, a non-negligible minority of the events display quite different behavior. We note that this correspondence rate may be affected by the fact that some CMEs in the dataset are not full halos. Therefore, their orientations may not be accurately determined.

Another feasible interpretation of these findings was discussed in Yurchyshyn *et al.* [45, 49]: the main axis of a CME may rotate as it expands into the interplanetary space. It could occur as a result of interaction between the ejecta and the heliospheric magnetic field due to coronal viscosity [50, 51]. The rotation may also occur because an erupting twisted flux is subject to kink or torus instability, so that its loop top could rotate as it evolves [52, 53, 54, 55]. Lynch *et al.* [56] and Gibson and Fan [57, also private communication] proposed that an expanding flux rope can reconnect with the surrounding fields and the footpoints of the erupting fields can be displaced, so the erupting flux rope may change its orientation.

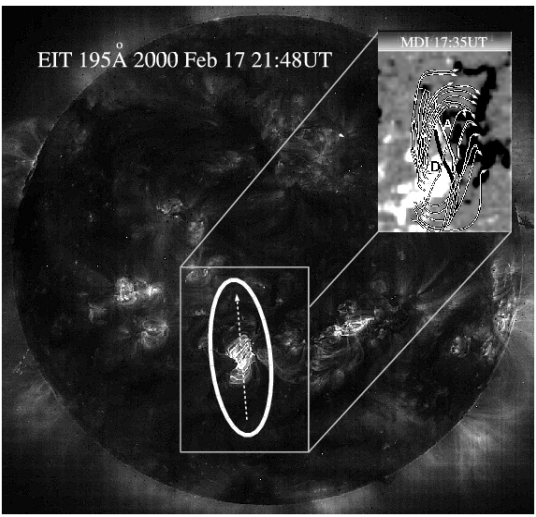

*Fig. 3. EIT 195Å image taken on 17 February 2000 at 2148 UT. The flare arcade associated with an eruption is inside the white ellipse that indicates the orientation of the corresponding CME. The white arrow inside the ellipse shows a northwardly directed axial field in the EIT arcade, determined from an Michelson Doppler Imager (MDI, [79]) magnetogram: from positive (white) to negative (black). The "S" shaped arcade indicates a right handed (positive helicity) magnetic structure. The field direction in the corresponding CME was ascribed, based on the EIT arcade and the acute angle method.*

The above possibilities will be addressed in the following sections. In order to address the above uncertainties and estimate the reliability of the above findings, a larger statistical study is needed, which would involve more magnetic parameters.

**2.2. *Orientation of CME elongations and direction of the axial field in EIT post eruption arcades***

In this section we will compare orientations of Earth-directed CMEs and the corresponding PEAs. Tripathi *et al.* [58] analyzed association between 236 PEAs observed by the EIT instrument and CMEs and found a nearly one to one correspondence between them. Based on this result, PEAs can be considered as reliable tracers of CME events. For each PEA event these authors determined its heliographic position and the length along its axis, which was overlying magnetic polarity inversion lines, when traced back to the Michelson Doppler Imager (MDI, [59]) synoptic charts of the photospheric magnetic field. The heliographic length of the PEAs was found to be in the range of 2 to 40 degrees, with an average of 15 degrees. No error analysis was performed in this study since the primary focus was the range of values and a trend of parameters.

The data set presented in [58] allowed us to determine the tilt (orientation) of these PEAs relative to the solar east-west line. This tilt was measured in degrees in clockwise direction starting from the east. We estimated that in most cases the PEA angles were measured with an accuracy of ±10 deg, although in 6 events the absolute error was estimated to be 90 deg.

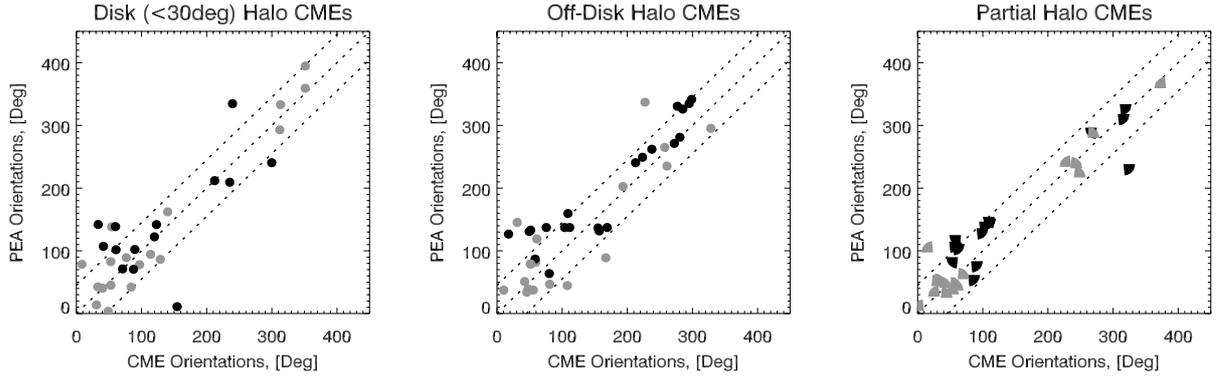

*Fig. 4. CME directional angles plotted versus those of PEAs. The panels show (from left to right) data points for disk halo CMEs that originated within a 30 deg circle centered at the disk center, off-disk halo CMEs and partial CMEs. Black (gray) symbols represent events launched from southern (northern) hemisphere. The partial halo events are plotted with pie segments, whose orientation indicates the solar disk quadrant where they originated.*

The list of Tripathi et al. [58] includes various events and not all of them are full and partial halos. For our purposes we needed to modify this list and discard all events associated with narrow and limb CMEs. Therefore, we were left with 101 PEA-CME events. For each full or partial halo CMEs from the list we determined its orientation as described above.

Since a PEA is a magnetic structure with a well defined axial field, we can take into account the PEA's axial field and assign direction to the orientation angles. The axial field and twist (helicity sign) in PEAs can be determined from solar data based on i) direct calculations of the predominant current helicity [60, 61, 62]; ii) force free field modelling [63]; iii) magnetic orientation angles [64]; and iv) visual inspection of the loop pattern seen in chromospheric and coronal images such as S-shaped sigmoids [10, 12] and/or dextral and sinistral filaments [65].

To compare these ambiguity resolved directions of PEAs with CME orientation angles, we need to also assign direction to the CME orientations. To do so we use the magnetic field information inferred from the associated PEA. The CME ambiguity resolution is based on the assumption that the direction of the axial field and twist in a flux rope CME corresponds to those of the EIT/Hα flare arcade associated with the eruption (see more details in Figure 3). We will, therefore, assume that the axial field of the PEA makes an acute angle with that of the CME.

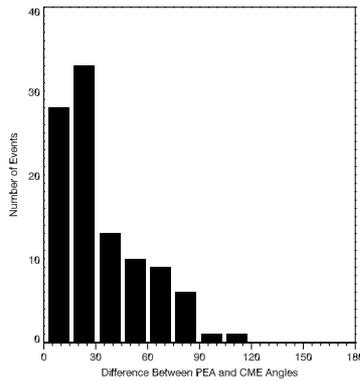

*Fig. 5. Distribution of angle difference between CME and PEA orientations. For 74 out of 101 events (73%) the difference is less than 45 deg.*

In Figure 4 we plot CME directional angles versus those of PEAs. The panels represent (from left to right) disk halo CMEs, off-disk halo CMEs and partial CMEs. First note that in all three cases the data points are mainly clustered near the bisector. In Figure 5 we show the distribution of PEA-CME difference angles plotted for all events. As it follows from the distribution plot, for the overwhelming majority of events (72%) the difference between the angles is less that 45 deg. It is worth emphasizing here that this holds true for both the partial halos and the halos that originate away from the disk center, where the projection effect is expected to play a significant role.

The data presented in Figures 4 and 5 indicate that, on average, the elongated halos are oriented along the axis of the corresponding PEAs. In addition, the elliptical shape of a halo CME may indeed bear information on the geometry of the underlying flux rope. These results also indicate that 64% correspondence between CME and MC orientations discussed in the previous section seems not to be caused by the projection effect in CME imaging, but is rather a result of CME evolution in the interplanetary space. In the next section we will address the relationship between solar and interplanetary magnetic fields by invoking magnetic field data at 2-3 solar radii above the solar surface.

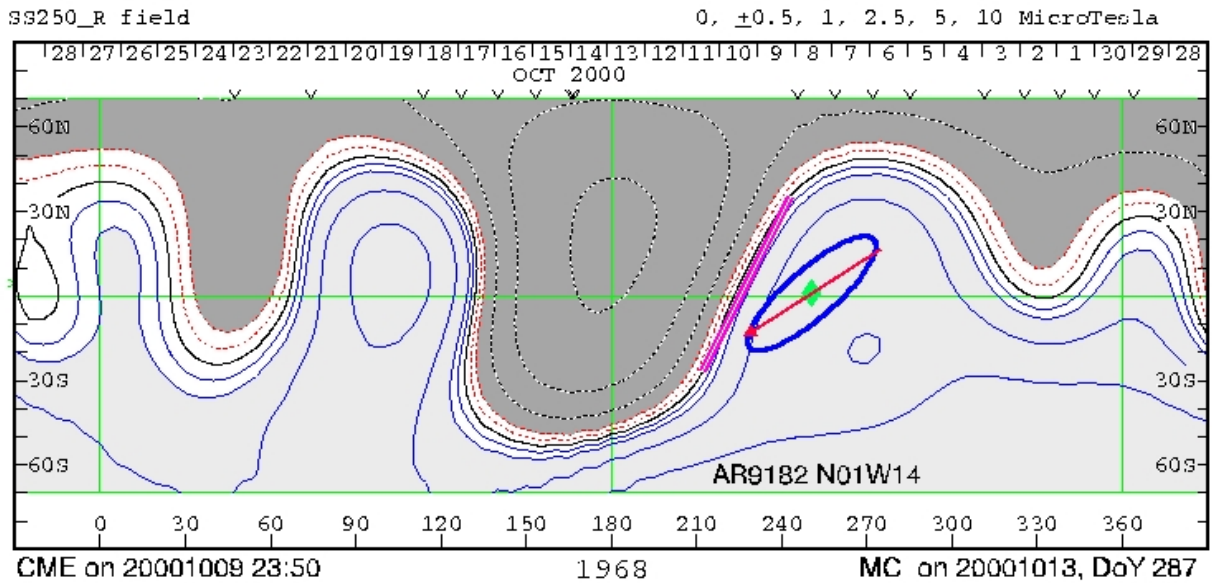

*Fig. 6. Wilcox Solar Observatory coronal field map calculated for the height of 2.5 solar radii for CR 1968. Dashed red contours (dark grey) represent negative fields, blue contours (light grey) – positive fields. Thick solid line is the coronal neutral line (CNL). A red line segment shows the estimated orientation of the CNL near the site of eruption on October 13, 2000. The tilt of the CNL was measured in the CW direction from the east. The ellipse, centered at the location of the CME source (diamond) indicates the orientation of a halo CME, while the red arrow shows the axial field direction of the corresponding magnetic cloud.*

## 3. Directional parameters of CMEs, coronal neutral line and magnetic clouds

It is generally believed that the wavy and spiral heliospheric current sheet [HCS, 66, 67] interacts with solar ejecta: whenever the ejecta are moving at a faster speed than the upstream plasma there must be an upstream influence. About half of these events drive upstream shocks, testifying to the fact of their "superior" speed in general [68, 69]. Zhao and Hoeksema [70] reported that the HCS may be blown out by a CME but shortly reforms near the same location.

Crooker *et al*. [71] suggested that the base of the HCS, which may be observed as a coronal neutral line (CNL), may often include multiple helmet streamers and that most CMEs might then be spatially associated with it. Subramanian *et al.* [72] estimated that about 65% of CMEs are associated with streamers. Moreover, the HCS is considered to be a conduit for CMEs [71] and the angular position (orientation) of the ejecta may, in general, be affected by its large-scale structure. When the current sheet is not aligned with the ejecta's plane, it is expected that the CME may displace the heliospheric magnetic field as it expands in the heliosphere [73]. As the magnetic field is distorted, it drapes around the ejecta and causes its deflection in longitude and/or azimuth. Since CMEs disrupt the global heliospheric magnetic field and the current sheet within it, it is natural to expect that CMEs themselves can be influenced, to some degree, by the heliospheric structures. If this is so, then the interaction between CMEs and the heliospheric magnetic field may explain the present difficulty in relating solar and interplanetary magnetic fields. However, details of this interaction are not yet studied well and the scale of the effect that the heliosphere may have on CMEs is largely unknown. Thus, the footprint of the photospheric magnetic fields associated with a CME may weaken and/or disappear due to changes in magnetic connectivity of the CME's field as suggested in Gibson and Fan [57], and/or changes in ejecta's shapes, structures, and dynamics caused by the interaction with ambient solar wind [74, 75, 76].

### 3.1. *Orientation of the coronal neutral line as measured from coronal field maps*

The background in Figure 6 is the Wilcox Solar Observatory (WSO) coronal magnetic field map showing the polarity distribution (light and dark gray) during Carrington rotation (CR) 1968. The black solid line represents the major CNL. This map was calculated from a synoptic photospheric field map with a potential field model [77, 78]. The green diamond in Figure 6 indicates the location of the CME source region relative to the CNL, i.e., on the day when it crossed the central meridian. The averaged tilt of the neutral line near the eruption site was measured (in clockwise direction) as the tilt of a thick line segment centered on the point closest to the eruption site.

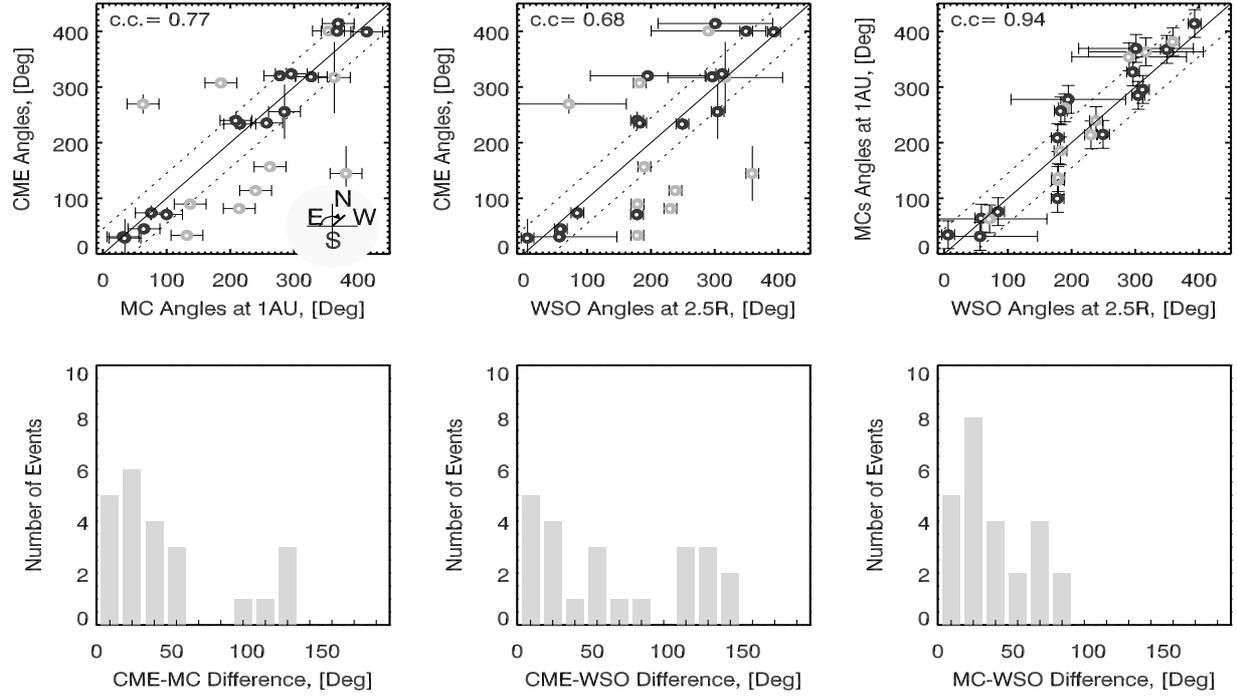

*Fig. 7. Top row, left: CME orientation angles plotted versus the axial field directions in MCs. The black symbols indicate those events that showed a good correspondence (difference < 45 deg) between the two directional angles, while the gray symbols are the events that display no correspondence (difference > 45 deg). Open error bars show standard deviation of the CME measurements, while closed bars are the absolute error of the MC data. Two dashed lines indicate ±45 degree interval around the bisector. Middle: CME orientation angles versus the orientation of the CNL near the eruption site. The black/gray symbols in this plot indicate the same events as in the left panel. Note that the majority of black symbols are also located at or near the bisector. Right: Directions of MC axial fields versus the orientation of the CNL. The black symbols indicate the same events as in the left panel. The majority of black and gray symbols in this plot are located at or near the bisector. Bottom row: Distribution of the angle difference between CME and MC (left), CME and CNL (middle) and MC and CNL (right) orientations. The bin size is 15 deg. The distribution of CNL-MC differences is very tight: for the majority of events (74%) the difference does not exceed 45 deg.*

In order to compare directional angles of CMEs and MCs with the tilt of the CNL, which has 180-deg ambiguity in it, we needed to assign the direction to the CNL orientation. This was done in a way similar to that for CMEs: by requiring that the CNL directional angle makes an acute angle with the axial field of the corresponding MC. Our choice of MCs as a reference is justified as follows: i) both MCs and CNL are low-order, large-scale heliospheric structures, as opposed to the PEAs that represent high-order solar surface fields and ii) it is not necessary that each PEA is formed under the streamer belt; therefore, the PEA orientation may not always be related to the CNL.

## 3.2. *Magnetic cloud parameters*

In a statistical study that involves solar surface activity, coronal eruptions and interplanetary disturbances, it is crucial to select events with reliably identified association between solar surface phenomena, CMEs and MCs. As soon as MCs are involved in the selection process, the number of available and suitable events sharply decreases so that the next statistical study is based so far on only 25 events.

For each event, the MC orientation angle was obtained by averaging MC orientations produced by different MC fitting routines [see Table I in 45]. The MC orientation angle is the direction angle of the projected MC flux rope axis onto the GSE *yz* plane, measured in the clock-wise direction from the positive *y*-axis. Current methodology and techniques allowed us to determine the MC axis position with accuracy no better than 20 deg.

### 3.3. *Results*

In Figure 7 (top row) we present relationships between CME, CNL and MC orientations. There are two groups of events presented with black and gray symbols. These events are separated as follows. In the upper left panel, black symbols indicate the 15 (out of 25) events that showed a good correspondence between the CMEs and MCs directions (angle difference < 45 deg). The gray symbols are events with the difference angle exceeding 45 deg. In the middle panel CME directions are compared to the CNL tilt. The black symbols here are also mainly clustered around the bisector. All "gray" events but two are also located outside the ±45 deg range centered on the bisector. Thus, the data presented in these two panels seem to suggest that those erupting flux ropes that where initially aligned with the CNL (streamer belt) at the early stage of eruption (black dots), appear to remain so when they reach the Earth, and thus that their orientations match those of the MCs.

In upper right panel of Figure 7, MC axis directions are plotted against the tilt of the CNL near the eruption site. This panel displays a high correlation between the parameters with the data points tightly clustered around the bisector. The lower panels in Figure 7 show corresponding distribution of difference angle. The figure indicates that the MCs in our data set tend to be aligned along the coronal neutral line. Considering that the CNL is the base of the heliospheric current sheet, it ultimately means that the MCs tend to be embedded in the current sheet.

### 4. Conclusions and Discussion

We have reviewed the relationships between CMEs, EUV-PEAs and the coronal neutral lines, associated with global magnetic field and magnetic clouds near the Earth. We find that i) there is a good correlation between the directions of the axial field in PEAs and the elongations of halo CMEs; ii) there is a good correlation between the orientations of the CNL and MCs: 80% of analysed events display the difference between the CNL and MC orientations to be less than 45 deg; iii) the majority of the events that had PEA and MCs similarly oriented also had the CNL co-aligned with them; iv) those events that showed no agreement between the PEA and MC orientations, had their MCs aligned with the CNL only.

As we mentioned earlier, observations and theoretical works indicate that the coronal ejecta may evolve substantially as it expands out into the heliosphere and interacts with heliospheric and solar wind magnetic fields. CMEs might have a tendency not only to be deflected toward the heliomagnetic equator and channeled into the HCS [70, 71, 79, 80], but also to be locally aligned with the heliospheric current sheet. This inference, based on a detailed study of 25 events is in agreement with the earlier reports that i) MCs oriented between ±30 deg, tend to be detected more frequently [25] and ii) during solar minimum (maximum) bipolar (unipolar) MCs dominate [79]. Note that bipolar (unipolar) MCs are nearly parallel (perpendicular) to the ecliptic.

An example that illustrates the suggested CME rotation is presented in Figure 8. We used the Potential Field Solar Surface [PFSS, 77] coronal field model results for CR 2006 obtained at Community Coordinated Model Center (CCMC). The 14 August 2003 eruption was relatively slow (~400km/s) and associated with a weak C3.6 flare in NOAA AR 10431. The left panels in Figure 8 show coronal field maps at 1.6 solar radii and the halo CME (ellipse), while the right panels are coronal field map at 2.5 solar radii and the corresponding MC is indicated with the cylinder. This CME was associated with the streamer belt and its elongation initially matched the local tilt of the neutral line at 1.6 solar radii. However, further out from the Sun, the neutral line changed its orientation, which is evident from the 2.6 solar radii map. It is quite possible that the associated CME rotated too, so that the corresponding MC was well aligned with the coronal neutral line. This is in agreement with Krall *et al*. [43] who reported that the 28 October 2003 CME smoothly rotated by about 50 deg until its final orientation closely matched the MC position.

Figure 8 also gives a hint for a possible interpretation of this event. The black dotted line in panel (a) roughly indicates the location and shape of the CNL. As can be inferred from the figure, the erupting flux rope could be kinked at the beginning, so that the loop top could rotate as it expands/unkinks [56]. Gibson & Fan [57, also private communication] proposed that an expanding flux rope can reconnect with the surrounding fields so that the footpoints of the erupting fields can be displaced. Another equal possibility is that a CME, which is not aligned with the HCS, pushes apart the magnetic field lines on both sides of the current sheet, so that the enhanced magnetic pressure and Lorentz force both work against the expanding fields, gradually deforming the loop top so that it locally aligns itself with the current sheet. The higher the speed of the ejecta, the shorter is the interaction time, so that very fast CMEs may not be affected by the current sheet at all or affected in a substantially lesser degree. Of course, nearby coronal holes may affect the ejecta, too [76]. Some CMEs originate from unipolar regions [14, 76] and are, thus, associated with unipolar boundary layer. In this case the above explanation may not be valid since HCS is not present.

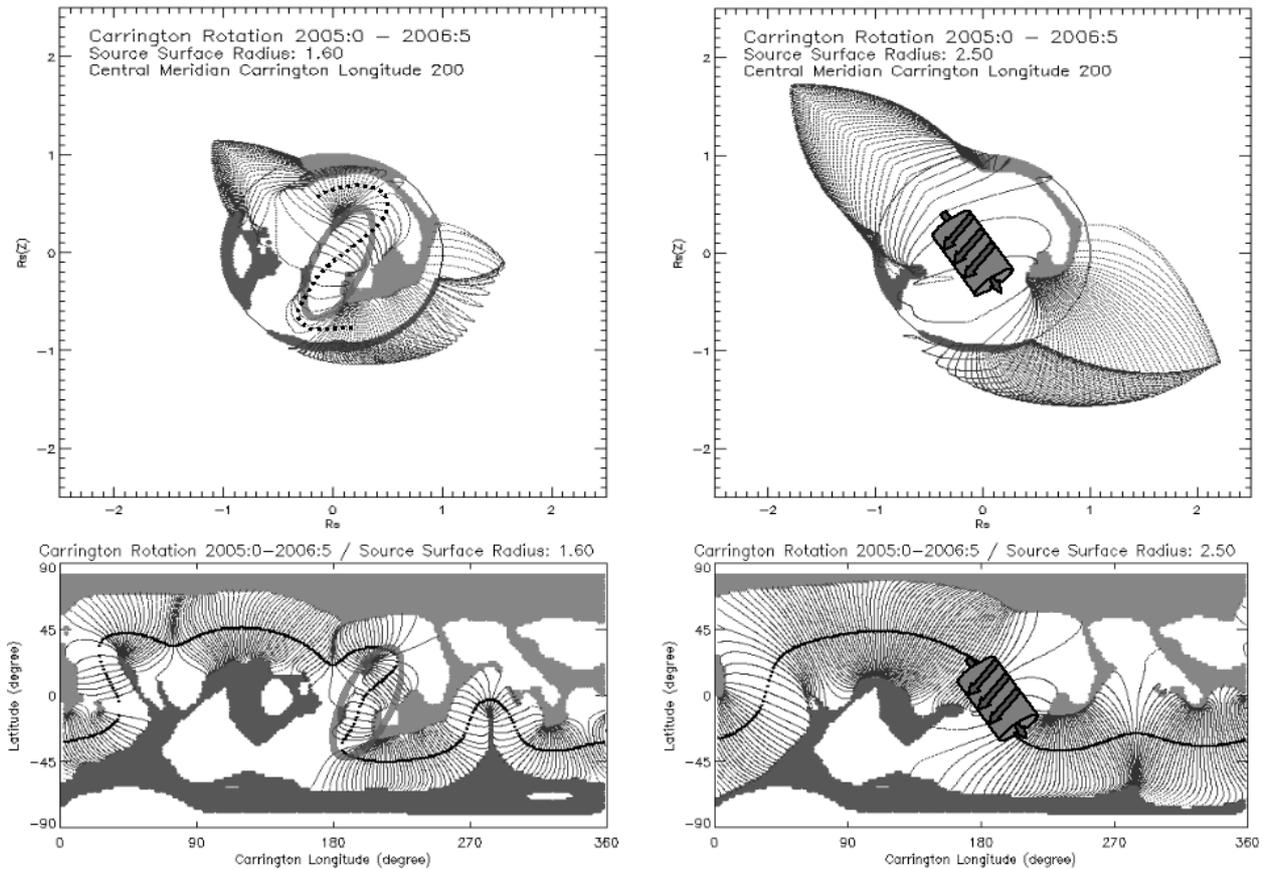

*Fig. 8. Coronal field maps calculated for CR2006 with Potential Field Solar Surface (PFSS) model run at Community Coordinated Model Center (CCMC). The left panels show maps for source surface at 1.6 solar radii; the right panels show maps at 2.5 radii. The thick black contour is the coronal neutral line. The oval represents the halo CME on 14 August 2003, which was aligned with the coronal neutral line at 1.6 solar radii. Magnetic topology has changed further outward from the solar surface so that the neutral line rotated by approx. 50 deg and was co-aligned with the magnetic cloud at 1AU (cylinder, lower right panel).*

## 5. Acknowledgements

The CME catalog is generated and maintained by the Center for Solar Physics and Space Weather, the Catholic University of America in cooperation with the Naval Research Laboratory, and NASA. We acknowledge the usage of the list of geomagnetic storms compiled during a *Living With a Star* Coordinated Data Analysis Workshops. SOHO is a project of international cooperation between ESA and NASA. Wilcox Solar Observatory data used in this study was obtained via the web site wso.stanford.edu, courtesy of J.T. Hoeksema. The work of VY was supported under NSF grant 0536921 and NASA ACE NNG0-4GJ51G and LWS TR&T NNG0-5GN34G grants. DT acknowledges support from STFC.